\documentstyle[12pt]{article}
\advance\textheight by 60pt
\advance\voffset by -40pt
\advance\textwidth by 50pt
\advance\oddsidemargin by -25pt
\advance\evensidemargin by -25pt

\def\single_space{\baselineskip 12pt plus 1pt minus 1pt}
\def\one_and_a_half_space{\baselineskip 19pt plus 1pt minus 1pt}
\def\double_spacesp{\baselineskip 25pt plus 2pt minus 2pt}

\newcommand{\leqnew}{\stackrel{<}{\!\ _{\sim}}}
\begin{document}
\begin{titlepage}
\begin{flushright}
{\bf July 1998} \\
{\bf Revised, September 1998} \\
{\bf Revised, October 1998}
\end{flushright}
\vskip 1.5cm
\double_spacesp
{\Large
{\bf
\begin{center}
Anatomy of a quantum `bounce'
\end{center}
}}
\vskip 1.0cm
\begin{center}
M.~A.~Doncheski \\
Department of Physics \\
The Pennsylvania State University\\
Mont Alto, PA 17237  USA \\
\vskip 0.1cm
and \\
\vskip 0.1cm
R. W. Robinett \\
Department of Physics\\
The Pennsylvania State University\\
University Park, PA 16802 USA \\
\end{center}
\vskip 1.0cm
\begin{abstract}
We discuss some of the properties of the `collision' of a quantum
mechanical wave packet with an infinitely high potential barrier,
focusing on novel aspects such as the detailed time-dependence of the
momentum-space probability density and the time variation of the
uncertainty principle product $\Delta x_t \cdot \Delta p_t$. We make
explicit use of Gaussian-like  wave packets in the analysis, but also
comment on other general forms.

\end{abstract}
\end{titlepage}
\double_spacesp

\begin{flushleft}
 {\large {\bf I.~Introduction}}
\end{flushleft}
\vskip 0.5cm

The use of wave packets to analyze the non-trivial 
time-dependence of quantum mechanical systems is one important aspect
of the study of the classical-quantum interface.  Popular
simulation packages \cite{styer} can help students visualize the
evolution of quantum states (as opposed to  time-independent
stationary state solutions \cite{liboff}, \cite{robinett_0},
\cite{ejp_h_atom}, \cite{robinett_1}) by allowing them to
continuously change parameters (such as the initial width of a
wave packet) to study what effect they have on the system under study.

A number of authors have considered various one-dimensional
quantum  mechanical problems in a wave packet approach studying 
transmission and reflection from square barriers
\cite{goldberg_1}, \cite{barrier_1}, \cite{barrier_1p}, \cite{edgar}
or linear potential steps \cite{barrier_2}, bound state wave packets
in single square wells in either position space \cite{greenman} or
momentum space \cite{segre}, in  double wells \cite{deutchman},
\cite{johnson},  or in systems of relevance  to solid state physics
\cite{periodic_1}, \cite{periodic_2}.  
Such examples of
wave packet behavior are also increasingly useful as  teaching tools
since the behavior of Coulomb wave
packets on circular \cite{brown} or elliptical \cite{nauenberg} orbits
are  being tested  experimentally on Rydberg atom systems \cite{yeazell},
\cite{review}.

Numerical methods for solving the time-dependent Schr\"odinger
equation have been discussed \cite{goldberg_1}, \cite{press} and, in
special cases, closed-form analytic results can be obtained by
use of the time-development operator
\cite{blinder}, \cite{robinett_acc},
$e^{-i\hat{H}t/\hbar} \psi(x,0) = \psi(x,t)$,
to solve the initial value problem.  Another approach is to
combine a large number of individual stationary state solutions 
for both unbound
\cite{merrill} and bound state problems \cite{greenman} to obtain wave
packets.   The most familiar example is the explicit calculation of
the Gaussian free-particle wave packet which is treated in the 
majority of elementary texts.  A simple variation is to consider a
particle, subject to the one-dimensional `infinite wall' potential
\cite{robinett_1} given by
\begin{equation}
V(x) = \left\{ \begin{array}{ll}
               0 & \mbox{for $x<0$} \\
               \infty& \mbox{for $x \geq 0$}
                                \end{array}
\right.
\label{wall}
\end{equation}
so that it is free for $x<0$, but a wavepacket impinging on the
`wall' at the origin will `bounce'.  Andrews \cite{andrews}
has shown how some of the most obvious aspects of the
`collision' process (namely the long-time development of the reflected
wave packet, interference effects during the `collision', etc.) 
can be understood by considering combinations of
free-particle `mirror' solutions, and we will use some of his
arguments. 

In this note  we will examine, in some detail,  the `bounce' of
free-particle wave packets from the infinite wall potential described
by Eqn.~(\ref{wall}), focusing on several other issues, namely the
behavior  of the  momentum-space wave packet solutions,
the widths of the position- and
momentum-space packets during the `bounce', and the uncertainty
principle product $\Delta x_t \cdot \Delta p_t$ as a function of time.
We will make extensive use of the free-particle Gaussian wave packet
in our discussion, but we also present results for other, more general
forms; for completeness sake, however, we very briefly review the
essentials of the Gaussian case.

A free-particle wave packet can be constructed, using any initial
momentum-space weighting function, $\phi(p,0)$, via
\begin{equation}
\psi(x,t) = \frac{1}{2\pi \hbar}
\int_{-\infty}^{+\infty}
e^{ipx/\hbar}e^{-ip^2t/2m\hbar}\,\phi(p,0)\,dp
\label{basic}
\end{equation}
to give a time-dependent position-space wavefunction, $\psi(x,t)$.
 The
momentum-space solution itself has a trivial time-dependence, namely
\begin{equation}
\phi(p,t) = \phi(p,0) e^{-ip^2t/2m\hbar}
\label{mom_time}
\end{equation}
For the case of a Gaussian momentum-distribution, here written in the
form 
\begin{equation}
\phi_1(p,0) = \sqrt{\frac{\alpha}{\sqrt{\pi}}}
e^{-\alpha^2 (p-p_0)^2/2}
e^{-ip x_0/\hbar}
\end{equation}
the necessary integral in Eqn.~(\ref{basic}) can be done to obtain the
well-known result
\begin{equation}
\psi_1(x,t) = \frac{1}{\sqrt{\alpha \hbar F \sqrt{\pi}}}
e^{i[p_0(x-x_0) - p_0^2t/2m]/\hbar}
e^{-(x-x_0-p_0t/m)^2/2\alpha^2 \hbar^2 F}
\label{gaussian_wave_packet}
\end{equation}
where $F = 1 + i t/t_0$ and $t_0 \equiv m\hbar \alpha^2$.
This solution describes a Gaussian position-space wave packet whose width
increases with time, characterized by  arbitrary initial  values of
$x_0$ and $p_0$.  The resulting position-space probability density is
\begin{equation}
P_{free}(x,t) = |\psi_1(x,t)|^2
= \frac{1}{\beta_t \sqrt{\pi}}
e^{-(x-x_0-p_0 t/m)^2/\beta_t^2}
\end{equation}
where
$\beta_t \equiv \alpha \hbar [1 + \left(t/t_0\right)^2]^{1/2}$
and various important expectation values are given by
\begin{equation}
\langle x \rangle_t
= x_0 + \frac{p_0 t}{m}
\quad
,
\quad
\Delta x_t = \frac{\alpha \hbar}{2} 
\sqrt{1 + \left(\frac{t}{t_0}\right)^2}
\quad
,
\quad
\langle p \rangle_t = p_0
\quad
,
\quad
\Delta p_t = \frac{1}{\sqrt{2} \alpha}
\end{equation}

For the infinite wall case, we also can obtain wave packet solutions
from Eqn.~(\ref{basic}) by substituting the appropriate plane wave
solutions
\begin{equation}
e^{ipx/\hbar}
\qquad
\longrightarrow
\qquad 
\left\{ \begin{array}{ll}
e^{ipx/\hbar} - e^{-ipx/\hbar}  & \mbox{for $x \leq 0$} \\
              0                 & \mbox{for $x \geq 0$}
                                \end{array}
\right.
\end{equation}
in the basic integral.  In this approach, the integrals must be
performed numerically. On the other hand, we
can also make use of the method of Andrews \cite{andrews} 
and use {\it any}
free-particle wave packet solution $\psi(x,t)$ via
\begin{equation}
\tilde{\psi}(x,t)  = \left\{ \begin{array}{ll}
               \psi(x,t) - \psi(-x,t) & \mbox{for $x \leq 0$} \\
               0 & \mbox{for $x \geq 0 $}
                                \end{array}
\right.
\label{andrews}
\end{equation}
which satisfies the Schr\"odinger equation for the potential in
Eqn.~(\ref{wall}) as well as the appropriate boundary condition at the
wall.  In either case, if the original free-particle wave packet is
properly normalized, the `bouncing' wavepackets will also be very
close to being normalized, provided they 
are initially
far enough from the wall so  that any contribution from the `tail' in
the $x>0$ region is negligible.  In either case, however, in order to
obtain the time-dependent momentum-space wavefunction, we must
numerically evaluate the Fourier transform
\begin{equation}
\phi(p,t) = \frac{1}{\sqrt{2\pi\hbar}}
\int_{-\infty}^{+\infty} e^{-ipx/\hbar}\, \psi(x,t)\,dx
\label{fourier}
\end{equation}

To illustrate the behavior of such a `bouncing' wave packet, we show
in Fig.~1 plots of the position- and momentum-space probability
densities for a Gaussian wavepacket for various times before and after
a collision.  We have used the following values in
numerical integrals:
\begin{equation}
\hbar = 1
\quad
,
\quad
m = 1
\quad
,
\quad
p_0 = 10
\quad
,
\quad
x_0 = -10
\quad
,
\quad
\alpha = 1
\label{standard_values}
\end{equation}
With these values,  the spreading time is $t_0 = 1$ and the time it
takes the packet to return to its initial starting point is $T=2t_0 =
2$, so that an appreciable amount of spreading is obvious.  In order
to see what features of such `collisions'  are specific to Gaussian
packets, in Fig.~2  we show the same plots, but for an initial
momentum-space amplitude given by a Lorentzian form, namely
\begin{equation}
\phi_2(p,0) = \sqrt{\frac{2\alpha}{\pi}}
\frac{1}{[\tilde{\alpha}^2(p-p_0)^2 + 1]}\,
e^{-ipx_0/\hbar}
\label{lorentzian}
\end{equation}
The corresponding initial position-space wavefunction is
\begin{equation}
\psi_2(x,0) =
\frac{1}{\sqrt{\tilde{\alpha}}\hbar}
e^{-|x-x_0|/\tilde{\alpha}\hbar}
\label{lorentz_fourier}
\end{equation}
but the further time-dependence can only be evaluated numerically
using Eqn.~(\ref{basic}). 
Using these two cases, we can make some general comments:
\newcounter{temp1}
\begin{list}
{(\roman{temp1})}{\usecounter{temp1}}
\item The  non-Gaussian position-space wave packet comes to approach
the Gaussian form more and more closely, as it evolves in time.  This
behavior is seen for a large number of other, single-humped initial
distributions \cite{robinett_comment}.  
\item The momentum-space probability density well after the collision
is related to the initial density by
$P_{after}(p,t) = P_{before}(-p,t)$
corresponding to the reversal of each momentum component during to the
collision.
\item At the moment of the collision, however, the momentum distribution is
{\it not} symmetric.  This is clearly due to the
fact that the high momentum components are preferentially in the
leading edge, and are the first to be reflected to negative values.
This also implies, as will be seen later, that the expectation value
$\langle p \rangle_t$ is slightly negative at $t = T_C$, the
collision time.   We note that
other `velocity effects' have been discussed  \cite{barrier_1},
\cite{barrier_1p} for  various kinds of wave packet scattering. 
\end{list} 
To focus on the details of the collision event, in Fig.~3 we plot
the time-dependent $|\psi(x,t)|^2$ and $|\phi(p,t)|^2$  (for the
Gaussian wave packet) for times nearer the actual `bounce', bracketing
$t=T_C$,  and we note some additional  aspects of the process:
\newcounter{temp2}
\begin{list}
{(\roman{temp2})}{\usecounter{temp2}}
\item The time-dependence of $\phi(p,t)$, which is non-trivial only
during the collision, is more clearly visible as is the
eventual return to `symmetry' of $|\phi(p,t)|^2$. 
\item The spread in the position-space probability density at the time
of the collision is substantially smaller than
$\Delta x_t$ either immediately before or after the collision.  We
will address this point below, using an analytical evaluation of
$\Delta x_{t = T_C}$.
\end{list}

In order to examine more of the differences between the purely
classical and quantum approaches to the collision of a point particle,
we  plot in Fig.~4 calculations  of the expectation values
$\langle x \rangle_t$ and $\langle p \rangle_t$,  which are easily
evaluated numerically. In this figure, we
show the expectation values of $x$ and $p$ (solid curves)  for the
bouncing wave packet,  as well as those for the free-particle wave
packet (dotted curves) for the standard set of parameters in
Eqn.~(\ref{standard_values}) except that we use the value of $\alpha =
0.5$; this value is chosen to make the spreading of the wave packets
more obvious since in this case $t_0 = m\hbar \alpha^2$  is much
smaller. We also indicate the `one sigma' limits given by $\langle x
\rangle_t \pm \Delta x_t$ and $\langle p \rangle_t \pm \Delta p_t$  as
dashed curves.

We first note that the guaranteed relationship between $\langle x
\rangle_t$ and $\langle p \rangle_t$, namely
$\langle p \rangle_t = m d\langle x \rangle_t/dt$,
is trivially observed long before and long after the collision,  while
the same qualitative connection between $\langle x \rangle_t$ and
$\langle p \rangle_t$ near $t = T_C$ is now  also apparent and
different than a purely classical `bounce' for a point particle which
would have a cusp (discontinuity) in $x(t)$ ($v(t)$) at the collision
time.  Finally, we can see that the position spread at
the collision time, $\Delta x _{t=T_C}$ is slightly smaller for the
`bouncing' wavepacket than for the free-particle packet with the same
initial parameters, while the momentum spread is much larger at $T_C$
than in the free-particle case.

In order to examine the `compression' of $\psi(x,T_C)$ and related
issues, we plot in Fig.~5 the values of $\Delta x_t$, $\Delta p_t$ and
the uncertainty principle product $\Delta x_t \cdot \Delta p_t$ (in
units of $\hbar$) for a range of values of $\alpha$,  but keeping the
other parameters fixed as in Eqn.~(\ref{standard_values}).  In each
case we see that $\Delta x_{t=T_C}$ is indeed smaller than its value
for the free-particle wave packet and in the cases where $\alpha \geq
1$, it is even smaller than it's original spread, $\Delta x_{t = 0}$.
This effect is perhaps intuitively obvious as the high momentum
components are reflected first, while the low momentum pieces `pile
up', leaving the position-space wave packet temporarily narrower.
This is not a violation of the $x-p$ uncertainty
principle as many other cases of such behavior are known; for example,
similar effects are seen 
in explicit constructions of wave packet solutions \cite{saxon} or more
simply in the direct  examination of the time-dependence of the
uncertainties in $x$ and $p$  in complete generality \cite{styer_1},
for the harmonic oscillator potential.  In our case, the fact that
$\Delta p_t$ does indeed increase  during the collision is even more
obvious, especially from the time-dependence of $|\phi(p,t)|^2$ shown
in Figs.~1 and 2.   During the collision, instead of being dominated
by the intrinsic width of  a single $\phi(p,t)$ peak, 
$\Delta p$ is dominated by the
distance between the peaks.  
    
>From the explicit numerical  calculations used to generate Fig.~5(a),
we find  to an excellent approximation that the position-space spread
at the collision time is given by
\begin{equation}
\frac{\Delta x^{(wall)}_{T_C}}{\Delta x^{(free)}_{T_C}} \approx 0.60
\end{equation}
and we can make use of the more analytic approach followed by Andrews,
at least at $t = T_C$ where the expressions simplify dramatically, to
understand this effect quantitatively.  Using the explicit
$\psi_1(x,t)$ in Eqn.~(\ref{gaussian_wave_packet}) and the expression
in Eqn.~(\ref{andrews}), we can construct an excellent approximation
to the `bouncing' wave packet for the Gaussian case, namely
\begin{equation}
|\tilde{\psi}(x,T_C)|^2  = 
\frac{4}{\beta_{T_C} \sqrt{\pi}}
\sin^2\left(\frac{p_0 x}{\hbar}\right)
e^{-x^2/\beta_{T_C}^2}
\end{equation}
which is approximately normalized, and the error is exponentially
small for the parameters we use, namely $p_0 \beta_{T_C}/\hbar >> 1$.
For these values, the $\sin^2(p_0x /\hbar)$
variation can very  reasonably be replaced  by its average value of
$1/2$ and the resulting integrals performed exactly.  We then have, to
an excellent approximation,
\begin{equation}
\langle x \rangle_{T_C} = -\frac{\beta_{T_C}}{\sqrt{\pi}}
\qquad \quad
\mbox{and}
\qquad \quad
\langle x^2 \rangle_{T_C} = \frac{\beta^2_{T_C}}{2}
\end{equation}
so that
\begin{equation}
\Delta x_{T_C} = \beta_{T_C}\sqrt{\frac{1}{2} - \frac{1}{\pi}}
\qquad \quad
\mbox{or}
\qquad \quad 
\frac{\Delta x_{T_C}^{(wall)}}
{\Delta x_{T_C}^{(free)}} = \sqrt{\frac{\pi -2}{\pi}}
\approx 0.603
\end{equation}
all of which are observed numerically! 

A similar semi-analytic result can be obtained which describes the
expectation value of the momentum operator at the collision time,
namely $\langle \hat{p} \rangle_{T_C}$.  The values of $\langle p
\rangle_t$ required for Figs.~4 and 5 have been obtained numerically,
by using the momentum space probability density, but using the
wavefunction representation in Eqn.~(\ref{andrews}), we can also
obtain an explicit formula for $\langle \hat{p} \rangle_t$ for the
special case of  $t = T_C$ using the representation of $\hat{p}$ as a
differential operator acting on the position-space wavefunction.  For
the free-particle wave packet we naturally have
\begin{equation}
\langle \hat{p} \rangle_{t}
= \int_{-\infty}^{+\infty} \, \psi_1(x,t)^*\, \hat{p}\,
\psi_1(x,t)\,dx
= p_0
\end{equation}
and the evaluation is straightforward
and well-defined for all times.  In contrast to this case,
if we naively attempt to evaluate $\langle \tilde{\psi}| \hat{p}|
\tilde{\psi} \rangle$ in this way we find that the expectation values
are  not necessarily Hermitian due to the `asymmetry' caused by the
presence of the wall. If, however, we instead adopt the `symmetrized'
version (which reduces to the standard value for the
free-particle case)
\begin{equation}
\langle \hat{p} \rangle_{T_C} =
\frac{1}{2}
\int_{-\infty}^{\,0} 
\left[
\left(\hat{p}\tilde{\psi}(x,T_C)\right)^*\,\tilde{\psi}(x,T_C)
+
\tilde{\psi}^{*}(x,T_C) \left(\hat{p}\tilde{\psi}(x,T_C)\right)\right]\,dx
\end{equation}
we find that $\langle \hat{p} \rangle$ is guaranteed to be real.  Using this
trick, we find that the expectation value at the collision time is 
\begin{eqnarray}
\langle \hat{p} \rangle_{T_C}
& = &  - \left(\frac{4 \hbar}{\sqrt{\pi}}\frac{t}{t_0}\right)
\frac{1}{\beta_{T_C}^3}
\left[
\int_{-\infty}^{\,0} \, x \, \sin^2\left(\frac{p_0 x}{\hbar}\right)
\, e^{-x^2/\beta_{T_C}^2}\,dx \right] \\ \nonumber
& \approx & 
- \frac{1}{\alpha \sqrt{\pi}}
\left[\frac{T_C/t_0}{\sqrt{1+(T_C/t_0)^2}}\right]
\end{eqnarray}
where $\sin^2(p_0 x/\hbar)$ term is replaced  by its average
value of $1/2$.  This analytic approximation agrees with all of our
numerical calculations to the desired accuracy.

\newpage

\begin{flushleft}
{\Large {\bf 
Figure Captions}}
\end{flushleft}
\vskip 0.5cm
 
\begin{itemize}
\item[Fig.\thinspace 1.]  Plots of a Gaussian wave packet striking an
infinite wall (bold vertical line.)
  $|\psi(x,t)|^2$ versus $x$ is shown on the left,
while the corresponding $|\phi(p,t)|^2$ versus $p$ plot is shown on the
right.   The initial position $x_0$ is shown on the left, while the
values of the central momentum long before ($+p_0$) and long after
($-p_0$) the collision are indicated on the right.  The numerical
values used are those in Eqn.~(\ref{standard_values}).
\item[Fig.\thinspace 2.] Same as Fig.~1, but for a wave packet
described by an initial Lorentzian  (given by
Eqns.~(\ref{lorentzian}) and (\ref{lorentz_fourier})
) which then evolve in time.
\item[Fig.\thinspace 3.] Same as Fig.~1, but for times nearer the
actual  `collision' at $t \approx T_C$. Note that the momentum
distribution at the moment of collision is {\it not} symmetric.
The peak near $+p_0$  is skewed towards values
with  $p \leqnew +p_0$, while the feature near $-p_0$ is similarly
enhanced with values just  below $-p_0$.  
\item[Fig.\thinspace 4.] Plot of $\langle x \rangle_t$ (top) and
$\langle p \rangle_t$ (bottom) as a function of time over the same
time interval as shown in Fig.~1.  In order to emphasize the spreading
of the wave packet, we use $\alpha = 0.5$; otherwise the parameters
are as in Eqn.~(\ref{standard_values}).   Results are shown for the
bouncing (solid curves) and the free-particle (dotted curves) packets.
The dashed lines indicate the one standard deviation spreads in  each
case, calculated numerically. Note that the expectation value of
momentum at the collision time $t = T_C$ is slightly negative.

\item[Fig.\thinspace 5.] Plots of (a) $\Delta x_t$, (b) $\Delta p_t$
and (c) the uncertainty principle product $\Delta x_t \cdot \Delta
p_t$ (in units of $\hbar$) versus $t$ for various `bouncing'
wavepackets over the same time interval as in Fig.~1.  The various
cases corresponding  to $\alpha = 3,2,1,1/2,1/3$ are given by the
dash-dash-dot, dash, solid, dot-dash, and dotted curves.  Otherwise,
the standard set of parameters in Eqn.~(\ref{standard_values}) are
used.  All the Gaussian packets shown start with $\Delta x_{t} \cdot
\Delta p_t = \hbar/2$.
\end{itemize}


\newpage
 
\begin{thebibliography}{99}
%
\bibitem{styer} See, for example, Hiller~J~R, Johnston~I~D and
Styer~D~F 1995 {\it Quantum Mechanics Simulations: The Consortium for
Upper-Level Physics Software} (New York: Wiley).
%
\bibitem{liboff} Liboff~R~L 1991 {\it Introductory Quantum Mechanics}
(Reading: Addison-Wesley) 2nd edition pp.~555-556.
%
\bibitem{robinett_0} Robinett~R~W 1995  Quantum and classical probability
distributions for position and momentum  Am. J. Phys. {\bf 63} 823-832.
%
\bibitem{ejp_h_atom} Rowe~E~G~Peter 1987  The classical limit of quantum
mechanical hydrogenic radial distributions  Eur. J. Phys.
{\bf 8}, 81-87. 
%
\bibitem{robinett_1} Robinett~R~W 1997 {\it Quantum Mechanics: Classical
Results, Modern Systems, and Visualized Examples} (New York: Oxford University
Press).
%
\bibitem{goldberg_1} Goldberg~A,  Schey~H~M, and Schwartz~J~L 1967
 Computer-generated motion pictures of one-dimensional quantum-mechanical
transmission and reflection phenomena  Am. J. Phys. {\bf 35} 
177-186.
%
\bibitem{barrier_1}    Bramhall~M~H and Casper~B~M 1970
 Reflections on a wave packet approach to quantum mechanical barrier
penetration  Am. J. Phys. {\bf 38} 1136-1145. 
%
\bibitem{barrier_1p} Diu~B (1980) 
 Plane waves and wave packets in elementary
quantum mechanics problems  Eur. J. Phys. {\bf 1} 231-240. 
%
\bibitem{edgar} Edgar~A 1995 Reflection of wave packets from a quantum
well with a tunneling transmission resonance Am. J. Phys.
{\bf 63} 136-141.
%
\bibitem{barrier_2} Boleman~J~S and Haley~S~B 1975 
 More time-dependent
calculations for the Schr\"odinger equation  Am. J. Phys. {\bf 43}
270-271. 
%
\bibitem{greenman} Greenman~J~V 1972 Non-dispersive mirror 
wave packets Am. J. Phys. {\bf 40} 1193-1201.
%
\bibitem{segre} Segre~C~U and Sullivan~J~D 1976  Bound-state wave
packets  Am. J. Phys. {\bf 44} 729-732.
%
\bibitem{deutchman} Deutchman~P~A 1971 Tunneling between two square
wells -- Computer movie Am. J. Phys. {\bf 39} 952-954.
%
\bibitem{johnson} Johnson~E~A and Williams~H~Thomas 1981
Quantum solutions for a symmetric double square well
Am. J. Phys. {\bf 50} 239-243.
%
\bibitem{periodic_1} Hamilton~J~C, Schwartz~J~L,  and Bowers~W~A 
1972  Computer generated films for solid state physics  Am. J. Phys.
{\bf 40}, 1657-1972. 
%
\bibitem{periodic_2} Friedmann~G. and Little~W~A 1993
 A study of a
wave function of a particle striking a crystal interface ,
Am. J. Phys. {\bf 61}, 835-843. 
%
\bibitem{brown} Brown~L~S 1972 Classical Limit of the hydrogen atom
Am. J. Phys. {\bf 41} 525-530.
%
\bibitem{nauenberg} Nauenberg~M 1989 Quantum wave packets on Kepler
elliptic orbits Phys. Rev. {\bf A40} 1133-1136.
%
\bibitem{yeazell} Nauenberg~M, Stroud~C, and Yeazell~J 1994
The classical limit of an atom Sci. Am. {\bf 270} 44-49.
%
\bibitem{review} For a review, see 
Alber~G and Zoller~P 1991 Laser excitation of electronic
wave packets in Rydberg atoms Phys. Rep {\bf 199} 231-280.
%
\bibitem{press} See, e.g., 
Press~W~H, Flannery~B~P, Teukolsky~S~A, and Wetterling~W~T
{\it Numerical Recipes: The Art of Scientific Computing}
(Cambridge: Cambridge University Press).
%
\bibitem{blinder} Blinder~S~M 1968 Evolution of a Gaussian wave packet
Am. J. Phys. {\bf 36} 525-526.
%
\bibitem{robinett_acc} Robinett~R~W 1996 Quantum mechanical time-development
operator for the uniformly accelerated particle Am. J. Phys. {\bf 64}
803-808.
%
\bibitem{merrill} Merrill~J~R 1973  The propagation of quantum mechanical
wave packets  Am. J. Phys. {\bf 41} 1101-1103. 
%
\bibitem{andrews} Andrews~M 1998 Wave packets bouncing off walls
Am. J. Phys. {\bf 66} 252-254.
%
\bibitem{robinett_comment} See, e.g., Ref.~\cite{robinett_1} pp.~60-61.
%
\bibitem{saxon} Saxon~D~S 1968 Elementary Quantum Mechanics (New York:
McGraw-Hill) pp.~144-147.
%
\bibitem{styer_1} Styer~D~F 1989 The motion of wave packets through their
expectation values and uncertainties Am. J. Phys. {\bf 58} 742-744.
%
\end{thebibliography}
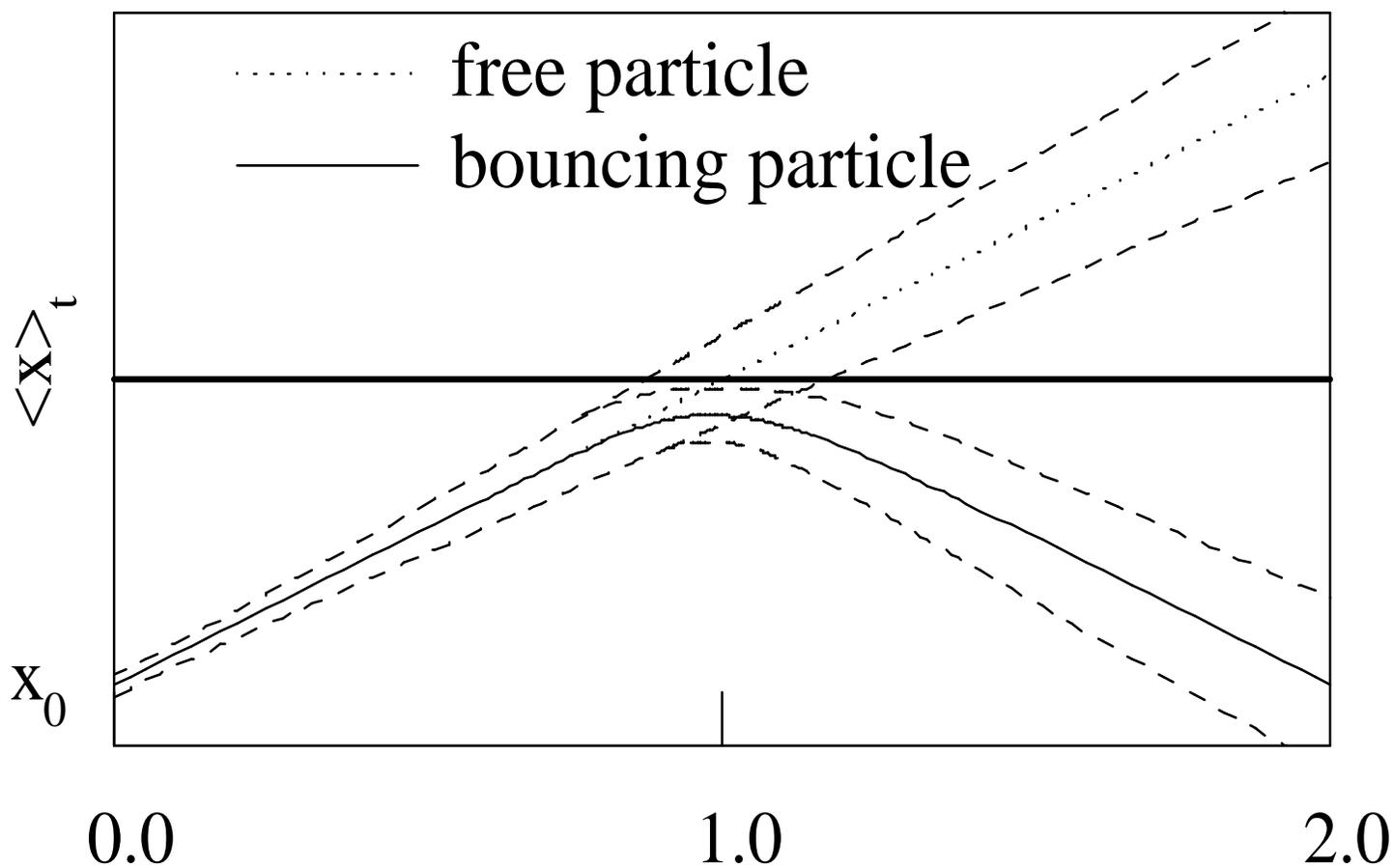
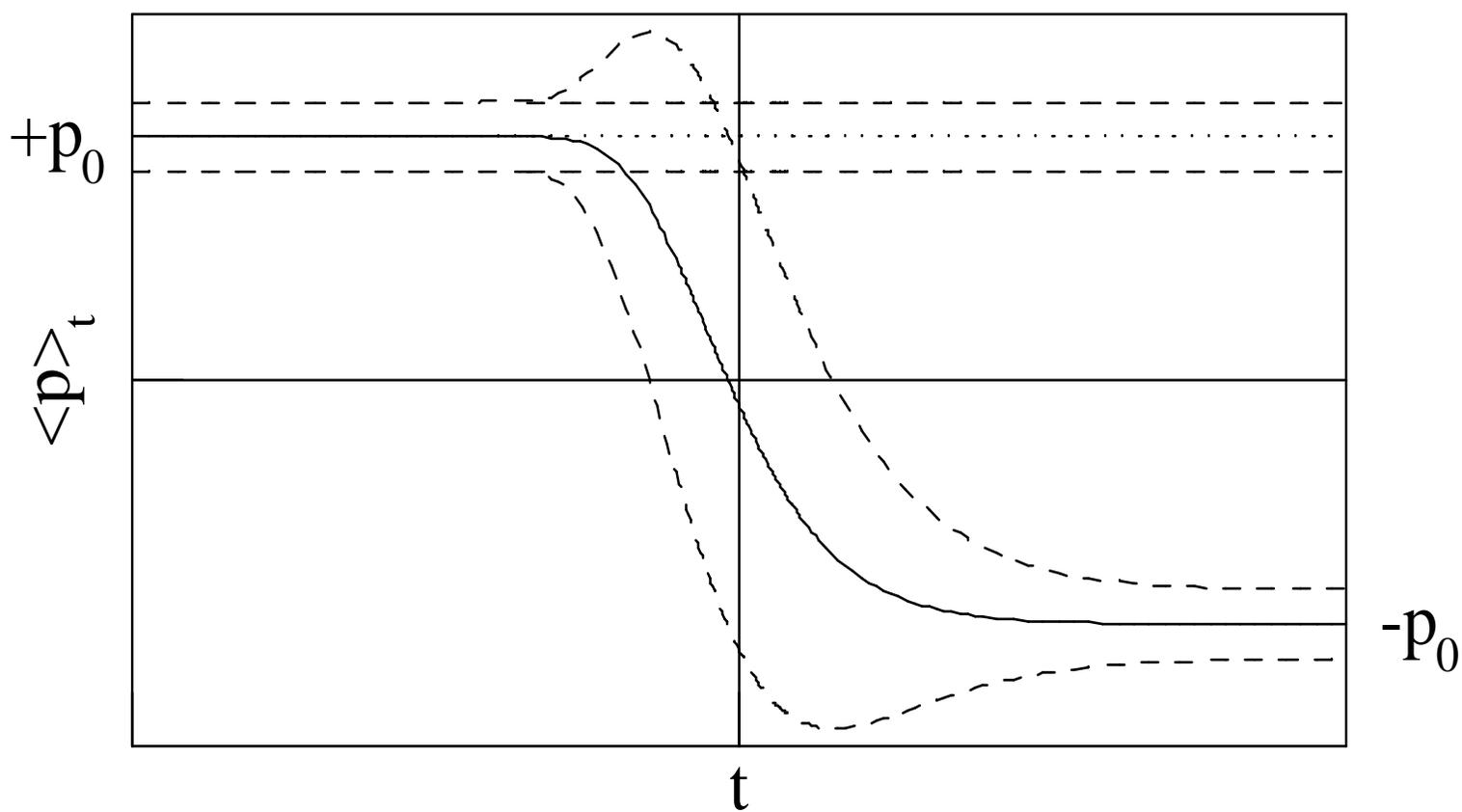
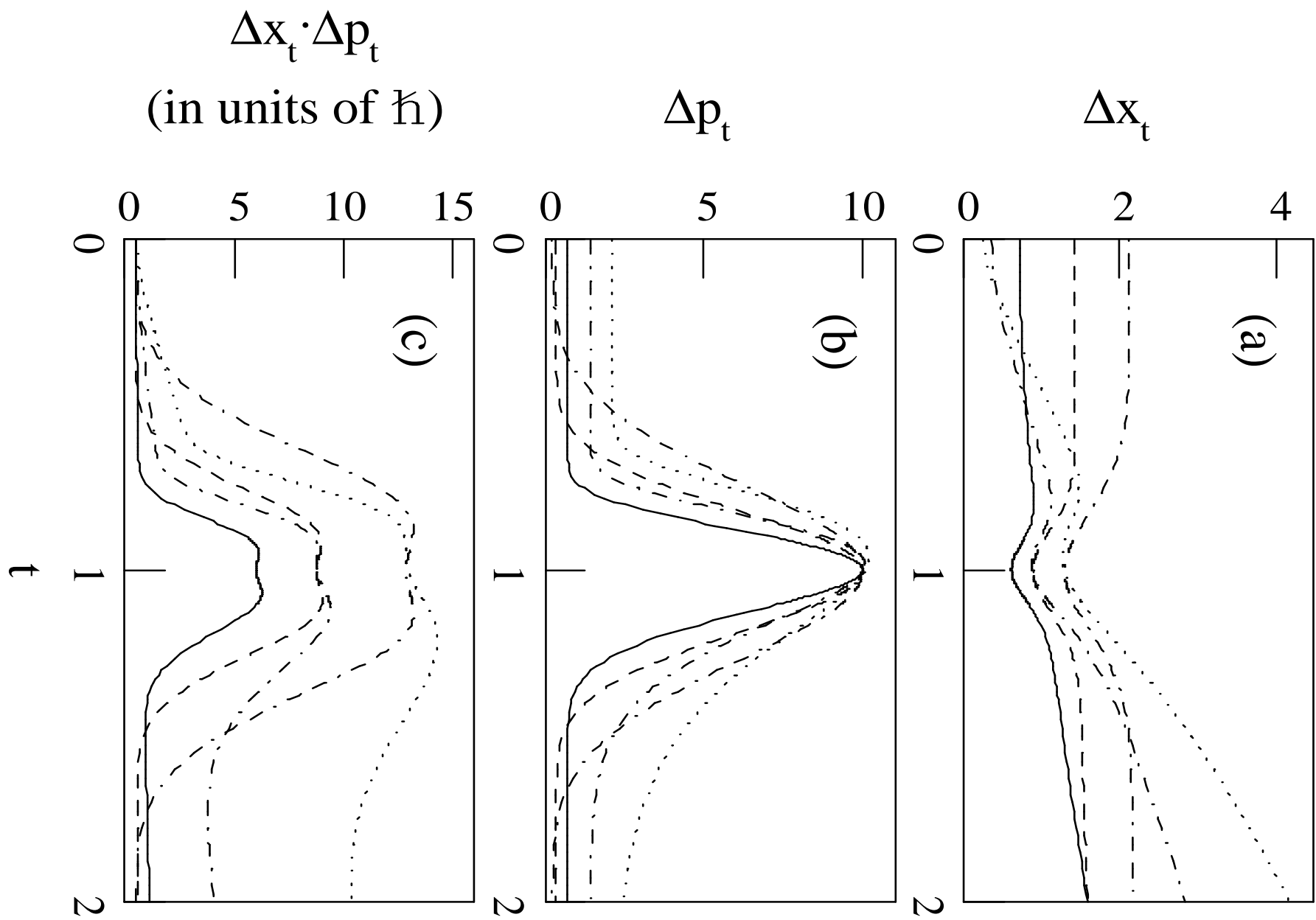
\end{document}